\documentclass[prd,aps,tightline,twocolumn,
nofootinbib,superscriptaddress]{revtex4}

\usepackage{amssymb}
\usepackage{amsmath}
\usepackage[dvips]{graphicx} 
\usepackage{latexsym}

\usepackage{color}

\definecolor{magenta}{cmyk}{0,1,0,0}

\begin{document}

\begin{flushleft}
STUPP-12-210, KEK-TH-1566
\end{flushleft}
\hfill \today

\title{Allowed slepton intergenerational mixing in light of light element abundances}

\author{Kazunori Kohri}
\affiliation{Theory Center, Institute of Particle and Nuclear Studies,
KEK (High Energy Accelerator Research Organization),
1-1 Oho, Tsukuba 305-0801, Japan}
\affiliation{The Graduate University for Advanced Studies (Sokendai), 1-1 Oho, Tsukuba
305-0801, Japan}

\author{Shingo Ohta}
\affiliation{Department of Physics, Saitama University, 
     Shimo-okubo, Sakura-ku, Saitama, 338-8570, Japan}

\author{Joe Sato}
\affiliation{Department of Physics, Saitama University, 
     Shimo-okubo, Sakura-ku, Saitama, 338-8570, Japan}

\author{Takashi Shimomura}
\affiliation{Department of Physics, Niigata University, Niigata, 950-2181, Japan}
\affiliation{Max-Planck-Institut f\"{u}r Kernphysik, Saupfercheckweg 1, 69117 Heidelberg, Germany}

\author{Masato Yamanaka}
\affiliation{Theory Center, Institute of Particle and Nuclear Studies,
KEK (High Energy Accelerator Research Organization),
1-1 Oho, Tsukuba 305-0801, Japan}


\begin{abstract}
We studied allowed region on the intergenerational mixing
parameters of sleptons from a viewpoint of big-bang
nucleosynthesis in a slepton-neutralino coannihilation
scenario. In  this scenario, $^7$Li and $^6$Li problems can be
solved by considering exotic reactions caused by bound-state
effects with a long-lived slepton.  Light element abundances  are
calculated as functions of the relic density and lifetime of the
slepton which considerably depend on the intergenerational 
mixing parameters. Compared
with observational light element abundances, we obtain allowed
regions on the intergenerational mixing. Ratio of
selectron component to  stau component, $c_e$, is allowed in
$2\times 10^{-11} \lesssim c_e \lesssim 2\times 10^{-9}$  with
solving both the $^7$Li and $^6$Li problems. Similarly, the ratio for
smuon, $c_{\mu}$,  is allowed in $
c_{\mu}
\lesssim 5\times 10^{-5}$
 for mass  difference between slepton and
neutralino, which is smaller than muon mass, and  $
c_{\mu} \lesssim 2\times 10^{-10}$ for the mass
difference in range between  muon mass and 125~MeV. We also
discuss collider signatures of the slepton decays. We find
characteristic double peaks in momentum distribution of event
number of the slepton decays with allowed mixing
parameters. Discoveries of the double peaks at future collider
experiments should confirm our scenario.
\end{abstract}

\maketitle

\section{Introduction}    

Long-lived Charged Massive Particles (CHAMP) are predicted in new
physics beyond the standard model~\cite{Fairbairn:2006gg}. Such
Long-lived CHAMPs  induce nonstandard nuclear reactions in big-bang
nucleosynthesis (BBN), and drastically change  light element
abundances~\cite{Pospelov:2006sc, Kohri:2006cn, Kaplinghat:2006qr,
Cyburt:2006uv, Steffen:2006wx,Bird:2007ge, Kawasaki:2007xb, Hamaguchi:2007mp,
Jittoh:2007fr, Jedamzik:2007cp, Pradler:2007is, Kawasaki:2008qe,
Jittoh:2008eq, Pospelov:2008ta, Kamimura:2008fx, Kusakabe:2008kf,
Bailly:2008yy, Bailly:2009pe, Kusakabe:2010cb, Jittoh:2011ni}. 
By comparing the theoretical predictions with observational data, we
can  obtain  constraints  on model parameters relevant with the
long-lived CHAMPs. Because the BBN is sensitive to lifetime from
$10^{-2}$~sec to $10^{12}$~sec, it should be  one of the best tools to
check an existence of the long-lived CHAMPs. \footnote{See also
Refs.~\cite{Jedamzik:2004er, Kawasaki:2004yh,  Kawasaki:2004qu,
Cumberbatch:2007me, Kohri:2008cf, Cyburt:2009pg, Pospelov:2010cw,
Kawasaki:2010yh} for long-lived neutral particles.}

The Minimal Supersymmetric Standard Model (MSSM) with R-parity conservation is one 
of the leading candidate providing long-lived CHAMPs. 
Several MSSM scenarios predict a long-lived stau as the next
lightest  supersymmetric particle (NLSP).  An example is the scenario
that the lightest  supersymmetric particle (LSP) is gravitino. In this
scenario, the interaction of the gravitino LSP with the stau NLSP is
suppressed by Planck scale, and hence the  stau has
longevity~\cite{Ellis:2003dn,Cerdeno:2005eu,
Steffen:2006hw,Cyburt:2006uv,Kawasaki:2007xb,Kawasaki:2008qe}.  
Another example is an axino LSP scenario, in which the decay rate of 
the stau NLSP is suppressed because of loop processes or a decay 
constant scale $F_{a}$~\cite{Covi:2004rb, Freitas:2011fx,Chun:2008rp}.
Among such scenarios, the most attractive one is a Bino-like
neutralino LSP scenario.  The longevity of the stau NLSP 
is brought by tight mass degeneracy of the stau and the neutralino LSP.

A remarkable feature of this mass-degenerate scenario is to be free
from the $^7$Li  problem \cite{Cyburt:2008kw}. The $^7$Li  problem is
a discrepancy between the observed abundance  of $^7$Li, e.g., ${\rm
Log}_{10}(^7\text{Li}/\text{H}) = -9.63 \pm
0.06$~\cite{Melendez:2004ni},  and a theoretical one predicted in the
standard BBN,  ${\rm Log}_{10}(^7\text{Li}/\text{H}) = -9.35 \pm
0.06$~\cite{Jittoh:2011ni}. (There is also a severer observational
limit which worsens the
fitting~\cite{Bonifacio:2006au}.)\footnote{This discrepancy can not be
solved even for corrections of corresponding cross sections of nuclear
reaction~\cite{Cyburt:2003ae,Angulo:2005mi}.  Even if we consider
nonstandard astrophysical models including diffusion
processes~\cite{Richard:2004pj,Korn:2006tv}, it would be difficult to
fit  all of data consistently~\cite{Lind:2009ta}. See also
~\cite{Iocco:2012rm} for a recent work related to light element
nucleosynthesis in accretion flows. }
In order to explain  the observed abundance of the dark matter by the
Bino-like neutralino LSP  based on a thermal relic scenario, the
so-called stau coannihilation mechanism is required to work
well~\cite{Griest:1990kh}. This mechanism requires of order or smaller than $1\%$ 
degeneracy in mass between the neutralino LSP and the stau NLSP. Due
to this  tight degeneracy, the stau can be long-lived 
\cite{Profumo:2004qt,  Jittoh:2005pq} and trigger exotic nuclear 
reactions at the BBN era.\footnote{See also
\cite{Sigurdson:2003vy,Hisano:2006cj,
Chuzhoy:2008zy,Borzumati:2008zz,Kohri:2009mi}
for another cosmological constraints on CHAMPs.}
One of such exotic reactions is an internal conversion process in a
bound state of the stau and $^7$Li ($^7$Be) nuclei~\cite{Jittoh:2007fr,
Bird:2007ge}. In the internal conversion process, $^7$Li ($^7$Be) can be 
destroyed into lighter nuclei and their abundances are fitted to the observational ones.
Therefore, in the mass-degenerate scenario, the observed abundances of 
the dark matter and all light elements are simultaneously realized.

It is further worth examining the BBN in the mass-degenerate scenario. 
In general, the MSSM has sources of intergenerational mixing, though
we had assumed the lepton flavor conservation in our previous works.
With the intergenerational mixing, the mass eigenstates are linear
combinations of  flavor eigenstates.  As we will discuss later in
detail, the intergenerational mixing reduces the number
density of the slepton via two types of processes. 
Success and failure of solving the $^7$Li problem strongly depends on 
the number density of the slepton. Furthermore, major and minor of 
modifications on the light element abundances are also determined by the 
number density of the slepton.
Thus, the modified abundances of the light elements can be predicted by 
the number density that is a function of mixing parameters. As a result, 
we can constrain the mixing parameters from the consistency between 
the modified and the observed abundances of the light elements.

The aim of this work is to predict the slepton mixing parameters in
the mass-degenerate scenario. In this scenario, the exotic nuclear
reactions catalyzed by the long-lived slepton induce both a  fusion
process~\cite{Pospelov:2006sc} and destruction processes through
internal  conversions for $^{7}$Li and
$^{7}$Be~\cite{Jittoh:2007fr,Bird:2007ge,Jittoh:2008eq,Jittoh:2010wh},
and spallations   for $^{4}$He~\cite{Jittoh:2011ni}. These processes
can modify the abundance of the  light elements. In fact, $^6$Li is
produced via the former process  after forming a bound state of the
slepton with $^4$He.  A fit by the observational abundance
$^6$Li/$^7$Li $= 0.046 \pm 0.022$~\cite{Asplund:2005yt}   gives an
allowed region in a parameter space of the mixing parameters. In
addition,  both deuterium (D) and tritium (T) (or $^{3}$He after its
decay) are produced nonthermally  by the latter processes.  By
adopting recent observational  constraints   D/H = $(2.80 \pm 0.20)
\times 10^{-5}$~\cite{Pettini:2008mq} and $^{3}$He/D $<$ 0.87 +
0.27~\cite{GG03},  an upper bound on the number density of the slepton
is obtained as a function of its lifetime. Thus, the upper bound on
the number density  results in a lower bounds on the mixing parameters
of the slepton.
On the other hand, a sufficient number density of the slepton at the
bound state formation with $^7$Li ($^7$Be) is required for solving the
$^7$Li problem. Then, an upper bound on the mixing parameters is obtained.
Hence, intriguingly, the mixing parameters of the slepton are predicted with
pinpoint accuracy from both above and below in light of the current 
observed light element abundances.

This long-lived slepton scenario can be also examined at terrestrial
experiments.  For example, due to the longevity longer than {\it``The
first  three minutes of the universe"},  a small number of the
long-lived slepton would  decay in detectors~\footnote{Collider
signatures of the  long-lived slepton in the neutralino LSP scenario
is studied in  case that its lifetime is shorter than the beginning of
BBN~\cite{Kaneko:2008re, Kaneko:2011qi} in collider experiments.
Measuring the lifetime and the mass difference between the slepton and
the neutralino LSP, the intergenerational mixing  will be
determined. Combining with the other works on collider signatures of
long-lived,  the models can be discriminated.}.
Some of them decay into the neutralino LSP and electron (or muon) via 
the intergenerational mixing. Combined with the predictions on the mixing 
parameters using the light element abundances, we find monochromatic 
and diffuse spectrum in momentum distributions of the final electron (or muon). 
These spectrum are characteristic signatures of the mass-degenerate scenarios, and 
therefore we can distinguish our scenario from others.

This paper is organized as follows. In the next section, we discuss the
relation between the number density of the long-lived slepton and the
intergenerational mixing in the mass-degenerate scenarios. Then, we
present a set of Boltzmann equations for calculating the number density
as a function of the mixing parameters.
In Sec.~\ref{sec:ana}, we analytically estimate bounds on the mixing parameters.  
Then, in Sec.~\ref{sec:result}, we show our numerical results. Allowed region on 
the intergenerational mixing parameter is shown from the viewpoint of observed 
light element abundances. In Sec.~\ref{sec:col}, applying the results, we 
discuss collider implications of this long-lived slepton. 
Finally, we summarize our results and discuss the prospects.

\section{Intergenerational mixing and number density of long-lived slepton}  
\label{sec:sle} 

Long-lived slepton leads some kinds of exotic nuclear reactions and
changes light element abundances. Predicted light element abundances
depend  on the number density of the slepton at the
BBN epoch. Interestingly, the  number density is quite sensitive to the
intergenerational mixing parameters  of the slepton in the
mass-degenerate scenarios.
Therefore, in order to constrain the intergenerational mixing, the
number density has to be accurately calculated as a function of the
intergenerational parameters.

A calculation of the number density of the slepton consists of two
steps. In the first step, the total number density of SUSY particles is
calculated at a time of their chemical decoupling.
After the chemical decoupling, the number density of the slepton
continues to evolve till the temperature $T \simeq \delta m$ via
scattering off the cosmic thermal background. Here $\delta m$ is the
mass difference between the slepton NLSP and the lightest neutralino
LSP. Furthermore the number of the slepton is reduced by its own 
natural decay.
Then, as the second step, the evolution of the number density is acquired by 
solving relevant Boltzmann equations implementing the scatterings and the decays. 
We explain the calculation of the number density, and then 
present a set of Boltzmann equations for the calculation.

\subsection{First step: Total number density of SUSY particles}  
\label{sec:tot} 

As the first step, we calculate the total number density of SUSY
particles. The total number density is equal to the relic number density
of the neutralino dark matter in the current universe because all SUSY
particles decay into the neutralino LSP in the end under the R-parity
conservation.
Calculation of the relic density of the neutralino dark matter in the
stau-neutralino coannihilation scenario has been developed so
far~\cite{Griest:1990kh, Edsjo:1997bg}.
Assuming the intergenerational mixing is not so large and the main
component of the slepton is stau, the calculation method is applicable
for our calculation.
That is, cross sections of coannihilation processes are firstly
calculated, and then we solve the Boltzmann equations for sum of the
number densities of SUSY particles.

After the chemical decoupling, the total number density of SUSYparticles 
remains the value at the freeze out. This value does not have correlation 
with the intergenerational mixing. This value is required as the initial 
condition for calculating the slepton number density as a function of 
the intergenerational mixing.

\subsection{Second step: Ratio of the number density between the 
slepton to the neutralino}  \label{sec:rat} 

\begin{figure} [!t]
\begin{center}
\includegraphics[width=8.5cm,clip]{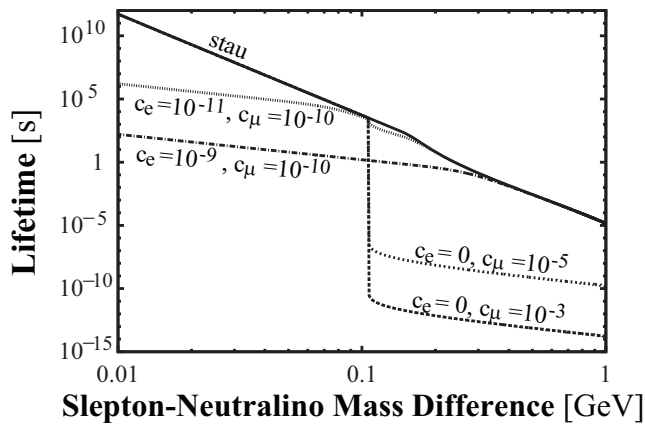}
\caption{Lifetime of the slepton as a function of $\delta m$ for 
$m_{\tilde l} = 350$GeV and $\sin\theta_e = 0.8$. Solid line 
is the lifetime of the pure stau, and others are the lifetime of 
slepton with each intergenerational mixing parameters. }
\label{fig:life}
\end{center}
\end{figure}

A key ingredient for constraining the mixing parameters is the evolution
of the number density of the slepton after the chemical decoupling of
SUSY particles.

Thermally-produced slepton decays according to its lifetime after the
chemical decoupling.  Hereafter we assume that the mass difference
$\delta m$ is smaller than the mass of tau lepton. In this situation,
the decay of slepton into the neutralino LSP and a tau lepton are
forbidden kinematically.  In this case the slepton becomes long-lived
\cite{Jittoh:2005pq}.

The decay of the slepton is categorized into two types. The first one is
3- or 4-body final state processes through the stau component;
\begin{equation}
\begin{split}
   &\tilde l^{\pm} \to \tilde \chi_1^0 + \nu_\tau + \pi^{\pm}, 
   \\
   &\tilde l^{\pm} \to \tilde \chi_1^0 + \nu_\tau + l^{\pm} + \nu_l 
   ~~ (l \in e, \mu). 
\label{dec_pro_tau}   
\end{split}     
\end{equation}
In the mass-degenerate scenario, the decay rates of these modes
considerably depend on $\delta m$ (see Fig.~\ref{fig:life} on this work
and Fig.2 in Ref.~\cite{Jittoh:2005pq}).
The second one is 2-body final state processes through the selectron or
smuon component;
\begin{equation}
\begin{split}
   &\tilde l^{\pm} \to \tilde \chi_1^0 + l^{\pm}
   ~~ (l \in e, \mu). 
\label{dec_pro_emu}   
\end{split}     
\end{equation}
The 2-body decay process are dominant when the mixing parameters are not
very small so that the decay rates are larger than those of 3- and 4-body decays.
On the contrary, when the intergenerational mixing is tiny, 3- or 4-body
decay processes dominate over 2-body decay processes in spite of a
tight kinematical suppression. The constraint on the mixing from these
decay processes must be therefore studied carefully as both functions of
$\delta m$ and the mixing parameters.

Exchange processes by scattering off the thermal background also play an
important role in constraining on the mixing parameters. Even after the
chemical decoupling of SUSY particles, the ratio of the number density
of the slepton to the neutralino continues to evolve via exchange
processes\footnote {Although other exchange processes exist, those
contributions are negligible due to weak interactions.},
\begin{equation}
\begin{split}
   &\tilde l^{\pm} + \gamma \leftrightarrow \tilde \chi_1^0 + \mu^{\pm}, ~~
   \tilde l^{\pm} + \gamma \leftrightarrow \tilde \chi_1^0 + \tau^{\pm}, 
   \\
   &\tilde \chi_1^0 + \gamma \leftrightarrow \tilde l^{\pm}  + l^{\mp} 
   ~~ (l \in e, \mu). 
\label{exc_pro}   
\end{split}     
\end{equation}
Note that the process $\tilde l \gamma \leftrightarrow \tilde \chi_1^0 e$ 
must not be included. The decay mode $\tilde l \to \tilde \chi_1^0 e$ is always 
open in the presence of the intergenerational mixing, and the process $\tilde l \gamma 
\leftrightarrow \tilde \chi_1^0 e$ should be counted as one of the corrective part for 
this decay (or inverse decay) mode with soft emission (or absorption). 
Thus, including the process $\tilde l \gamma \leftrightarrow \tilde\chi_1^0 e$ 
in addition to this decay mode results in a double counting. Similarly, when the 
mode $\tilde l \to \tilde \chi_1^0 \mu$ is open, then the corresponding exchange 
process $\tilde l \gamma \leftrightarrow \tilde \chi_1^0 \mu$ also must be 
taken out.

These exchange processes keep the slepton and the neutralino in kinematical 
equilibrium.  Due to this kinematical equilibrium, the number density ratio 
between the slepton and the neutralino maintain the Boltzmann distributions.  
The number density of the slepton for a fixed $\delta m$ therefore decreases as 
the temperature of the Universe decreases;
\begin{equation}
\begin{split}
    n_{\tilde l} =  \frac{n_{\tilde l}}{n_{\tilde \chi}} 
    \frac{n_{\tilde \chi}}{n_{\tilde \chi} + n_{\tilde l^+} 
    + n_{\tilde l^-}} N 
    = e^{- \delta m/T} 
    \frac{N}{2(1 + e^{- \delta m/T})},
\label{ratio}   
\end{split}     
\end{equation}
where $T$ is the temperature, and  $N$ is a sum of the densities of
the slepton and the neutralino that  are obtained in the first step
(see Sec.~\ref{sec:tot}). The kinematical equilibrium is broken when
the Hubble  expansion rate overwhelms the reaction rate of the exchange 
processes.  Hence it is important to know when the Hubble expansion rate 
overwhelms the reaction rate. We numerically solve a set of Boltzmann  
equations to obtain accurate resultant number density of the slepton.

\subsection{A set of Boltzmann equations for relic number density of the slepton}  
\label{sec:bol} 

Now we present a set of the Boltzmann equations to calculate the number density of 
the slepton as a function of the intergenerational mixing parameters.
The Boltzmann equations describing evolutions of the number densities of the 
negatively charged slepton $n_{\tilde l^-}$, the positively charged slepton $n_{\tilde l^+}$, and the 
neutralino $n_{\tilde \chi}$ are, 
\begin{equation}
\begin{split}
   &\frac{dn_{\tilde l^-}}{dt} + 3H n_{\tilde l^-} = 
   - \sum_{i, X, Y}
   \Bigl[ \langle \sigma' v \rangle_{i Y} n_{\tilde l^-} n_X^{eq} 
   - \langle \sigma' v \rangle_{\tilde l^- X} n_i n_Y^{eq} \Bigr]
   \\& \hspace{10mm}
   - \sum_{X, Y, ...} 
   \Bigl[ \langle \Gamma \rangle_{\tilde \chi X Y ...} n_{\tilde l^-} 
   - \langle \Gamma \rangle_{\tilde l^-} n_{\tilde \chi} 
   n_X^{eq} n_Y^{eq} ... \Bigr], 
\label{BE_l-}   
\end{split}     
\end{equation}
\begin{equation}
\begin{split}
   &\frac{dn_{\tilde l^+}}{dt} + 3H n_{\tilde l^+} = 
   - \sum_{i, X, Y}  
   \Bigl[ \langle \sigma' v \rangle_{i Y} n_{\tilde l^+} n_X^{eq} 
   - \langle \sigma' v \rangle_{\tilde l^+ X} n_i n_Y^{eq} \Bigr]
   \\& \hspace{10mm}
   - \sum_{X, Y, ...} 
   \Bigl[ \langle \Gamma \rangle_{\tilde \chi X Y ...} n_{\tilde l^+} 
   - \langle \Gamma \rangle_{\tilde l^+} n_{\tilde \chi} 
   n_X^{eq} n_Y^{eq} ... \Bigr], 
\label{BE_l+}   
\end{split}     
\end{equation}
\begin{equation}
\begin{split}
   &\frac{dn_{\tilde \chi}}{dt} + 3H n_{\tilde \chi} = 
   - \sum_{i, X, Y}  
   \Bigl[ \langle \sigma' v \rangle_{i Y} n_{\tilde \chi} n_X^{eq} 
   - \langle \sigma' v \rangle_{\tilde \chi X} n_i n_Y^{eq} \Bigr] 
   \\& \hspace{10mm}
   - \sum_{i, X, Y, ...} 
   \Bigl[ \langle \Gamma \rangle_{i} 
   n_{\tilde \chi} n_X^{eq} n_Y^{eq} ... 
   - \langle \Gamma \rangle_{\tilde \chi X Y ...} n_i   \Bigr],
\label{BE_chi}   
\end{split}     
\end{equation}
where $t$ is time and $H$ is the Hubble parameter. In the right hand side, $\langle \sigma' v
\rangle$ is the thermal averaged cross sections of the exchange
process in Eq.~\eqref{exc_pro}, and $\langle \Gamma \rangle$ is the
thermal averaged decay rate (or inverse decay rate). The
index $i$ represents relevant SUSY particles, and indices $X$ and $Y$
represent the SM particles, respectively.  Subscripts to both $\langle \sigma' v
\rangle$ and $\langle \Gamma \rangle$ denote the final states of each
process. 
In the equations, we assumed that the relevant SM
particles are thermalized\footnote{Vability of this
assumption is checked in Ref.~\cite{Jittoh:2010wh}. }.

In the presence of the intergenerational mixing, $\langle \sigma' v \rangle$ 
and $\langle \Gamma \rangle$ depend on the mixing parameters. Solving a set of 
the Boltzmann equations above, 
the number density of the slepton at the BBN epoch is 
obtained as a function of the mixing parameters. 
By using obtained number density, we  perform the BBN
calculations and predict light element abundances. Comparing them with
observational values, we can constrain  or predict the mixing
parameters.

\section{Analytical constraint on the intergenerational mixing}  
\label{sec:ana} 

Before showing our numerical results, we give analytical estimations for bounds on the 
mixing parameters. 
We expand the lightest slepton in terms of the  flavor eigenstates, 
$\tilde{e},~\tilde{\mu}$ and $\tilde{\tau}$ without the intergenerational mixing, 
\begin{equation}
\begin{split}
   \tilde{l}   &= \sum_{f=e,\mu,\tau}  c_f \tilde{f}, \\
     \tilde{f} &= \cos\theta_f \tilde{f}_L + \sin \theta_f  \tilde{f}_R,
\label{def:sle}   
\end{split}     
\end{equation}
where $\tilde{f}_L$ and $\tilde{f}_R$ are left-handed and right-handed
sleptons, and $\theta_f$ are the mixing angles of these sleptons,
respectively.  The coefficients $c_e$, $c_\mu$, and $c_\tau$ are the
mixing parameters. These parameters are in ranges from zero to unity and
satisfy the relation $c_e^2 + c_\mu^2 + c_\tau^2 = 1$.  In this work, we
assume $c_e$, $c_\mu$, and $c_\tau$ to be real and positive parameters
for simplicity.

These mixing parameters are constrained from above by two
requirements. The first requirement is the sufficient longevity of the
slepton, and the second one is the sufficient number density at the
moment of decoupling from the exchange processes.  For the first
requirement, the lifetime of the slepton must be longer than
$\mathcal{O}(10^3\text{s})$ to solve the $^7$Li problem. The slepton
with this lifetime can form a bound state with $^7$Li and $^7$Be and
destroy these nuclei via the internal conversion
processes~\cite{Jittoh:2007fr, Bird:2007ge}.
The lifetime is depicted as a function of $\delta m$ in Fig.~\ref{fig:life} for 
$m_{\tilde l} = 350$GeV and $\sin\theta_e = \sin\theta_\mu=0.8$. 
A curve labeled ``stau'' is the lifetime for a pure stau, and other curves are 
the lifetime for the slepton with the mixing parameters labeled near the curves. 
The lifetime varies by a few order of magnitudes as $\delta m$ and $c_{e,\mu}$ vary.
Due to the kinematical thresholds for the decays, dependence on both $c_e$ and 
$c_\mu$ is different for $\delta m > m_\mu$ or $\delta m < m_\mu$. 
Therefore, the lifetime should be studied in each $\delta m$ region to obtain bounds on the 
mixing parameters.

To obtain the bounds analytically, we consider two simple cases in
which either $c_e=0$ or $c_\mu=0$  for simplicity. For $c_\mu=0$ case,
the slepton decays into  an electron and the lightest neutralino
($2$-body decay).  Then the lifetime of the slepton is approximately
given by
\begin{align}
\tau_{\tilde{l}} (\tilde l \rightarrow \tilde{\chi}_1^0 + e) &\simeq 
\frac{8 \pi}{g^2 \tan^2 \theta_W} \frac{m_{\tilde{l}}}{(\delta m)^2}
                  \frac{1}{\cos^2 \theta_e + 4 \sin^2 \theta_e}
                  \frac{1}{c_e^2},
\end{align}
where $m_{\tilde \l}$ is the mass of the slepton. By inserting $\cos \theta_e = 0.6$ for 
a reference value, the lifetime is expressed as 
\begin{equation}
\begin{split}
   &\tau_{\tilde l} \ (\tilde l \to \tilde \chi _1^0 + e)  
   \simeq 1.34 \times 10^3 \nonumber 
   \biggl[ \frac{m_{\tilde \l}}{300\text{GeV}} \biggr] 
   \\& ~~ 
   \times 
   \biggl[ \frac{0.1\text{GeV}}{\delta m} \biggr]^2 
   \biggl[\frac{4-3 (0.6)^2}{4-3\cos ^2\theta _e}\biggr]
   \biggl[ \frac{10^{-10.5}}{c_e} \biggr]^2
   \text{s}. 
\label{app_lifetime_1} 
\end{split}     
\end{equation}
Requiring $\tau_{\tilde{l}} \ge \mathcal{O}(10^3\text{s})$, we obtain the bound on the 
selectron component, $c_e \lesssim 3.2 \times 10^{-11}$ for $\delta m = 0.1$ GeV and 
$m_{\tilde{l}} = 300$ GeV.
Similarly, for $c_e=0$ case, the lifetime is approximately given as 
\begin{equation}
\begin{split}
   &\tau_{\tilde l} \ (\tilde l \to \tilde \chi _1^0+ \mu) \simeq 
   10^3 
   \biggl[ \frac{10^{-10.5}}{c_\mu} \biggr]^2
   \biggl[ \frac{0.12\text{GeV}}{\delta m} \biggr]^2
   \\& ~~~~ \times 
   \biggl[ \frac{m_{\tilde l}}{300\text{GeV}} \biggr]
   \biggl( 1 - \frac{25}{36} 
   \biggl[ \frac{0.12\text{GeV}}{\delta m} \biggr]^2 \biggr)^{-1/2} 
   s.,
\label{life_mu_chi}   
\end{split}     
\end{equation}
where $\cos \theta_\mu = 0.6$ is used. From Eq.~\eqref{life_mu_chi},
the upper bound on $c_\mu$ is obtained as  $3.2 \times 10^{-11}$ for
$\delta m = 0.1$ GeV and $m_{\tilde{l}} = 300$ GeV.

We can derive another upper bounds on the mixing parameters from the second requirement 
that the number density should be sufficient at the decoupling time from the exchange 
processes to destroy the $^7$Li and $^7$Be nuclei.  
Omitting smuon component again, the exchange processes are $\tilde l e \leftrightarrow 
\tilde \chi_1^0 \gamma$ and $\tilde l \tau \leftrightarrow \tilde \chi_1^0 \gamma$ 
through selectron and stau component. 
The former exchange process keeps working even after the freeze out of the latter process. 
This is because densities of electron and photon are 
still quite large in thermal bath for $T \lesssim m_\tau$. 
Then, as explained in the previous section, the slepton number density
continues to decrease.
The reaction rate with the electron is proportional to $c_e^2$.
Hence the larger the mixing is, the more the number density
decreases.  The reaction rate of $\tilde l e \leftrightarrow \tilde
\chi_1^0 \gamma$  must be suppressed to ensure a sufficient number
density of the slepton.  The bound on $c_e$ is estimated by comparing
this reaction rate with that  of processes with tau.

In the absence of the intergenerational mixing or pure stau case, freeze out of the 
exchange processes ($\tilde l^{\pm}  \gamma \leftrightarrow \tilde \chi_1^0  
\tau^{\pm}$ and $\tilde \chi_1^0  \gamma \leftrightarrow \tilde l^{\pm} \tau^{\mp}$) 
occurs at $T \simeq 70$MeV, and is almost independent of both $\delta m$ and 
$m_{\tilde l}$~\cite{Jittoh:2010wh}. Then the number density of the slepton 
manages to be a sufficient amount.  This fact suggests that the reaction rate of 
$\tilde l e \leftrightarrow \tilde \chi_1^0 \gamma$ must be smaller than that of the 
processes in the absence of the intergenerational mixing at $T = 70$MeV. 
Parameterizing the cross section of this process as $\langle \sigma' v \rangle_e 
= c_e^2 \langle \sigma' v \rangle_\tau$, the ratio of these reaction rates at $T = 70$MeV 
is given by 
\begin{equation}
\begin{split}
   \frac{\langle \sigma' v \rangle_e Y_{\tilde l} Y_e}
  {\langle \sigma' v \rangle_\tau Y_{\tilde l} Y_\tau} 
  \simeq (1.08 \times 10^9) c_e^2, 
\label{ratio_of_reaction_rate}   
\end{split}     
\end{equation}
at $Y_{\tilde{l}}$ and $Y_{e,\tau}$ are the yield value of the slepton and electron/tau, respectively. 
The mixing parameter $c_e$ is therefore required to be $\lesssim 3 \times 10^{-5}$. 
For $c_e=0$ case, the reaction rate of the process $\tilde l \mu \leftrightarrow 
\tilde \chi_1^0 \gamma$ is compared with that without the mixing. Parameterizing the 
cross section of the process as $\langle \sigma' v \rangle_\mu 
= c_\mu^2 \langle \sigma' v \rangle_\tau$, the ratio of the reaction rates 
is given by
\begin{equation}
\begin{split}
   \frac{\langle \sigma' v \rangle_\mu Y_{\tilde l} Y_\mu}
  {\langle \sigma' v \rangle_\tau Y_{\tilde l} Y_\tau} 
  \simeq (9.93 \times 10^7) c_\mu^2, 
\label{ratio_of_reaction_rate2}   
\end{split}     
\end{equation}
at temperature $T = 70$MeV. From Eq.~\eqref{ratio_of_reaction_rate2}, 
$c_\mu$ must to be smaller than $10^{-4}$.

As we will see later, we can constrain $c_e$ from below
by the relic abundance of light elements. 
Thus we will get an allowed region for $c_e$.

\section{Numerical results}  \label{sec:result} 

We are now in position to numerically search for an allowed region of the 
slepton mixing parameters. First we see that the upper bounds on $c_e$ 
and $c_\mu$ obtained in the previous section are in a good agreement 
with numerical analysis. 
Then we numerically compute reaction networks of light elements
including  the exotic nuclear reactions, and find  parameter regions
as a function of  $\delta m$ allowed by the observational light
element abundances.

\subsection{Constraint on slepton intergenerational mixing}  \label{sec:inter} 

First we show bounds on the mixing parameter $c_\mu$ for $\delta m < m_\mu$.
Based on previous works~\cite{Jittoh:2007fr}, we fix $Y_{\tilde l} = 2 \times 
10^{-13}$ for the sufficient number density to solve the $^7$Li problem.

\begin{figure} [!t]
\begin{center}
\includegraphics[width=8.6cm,clip]{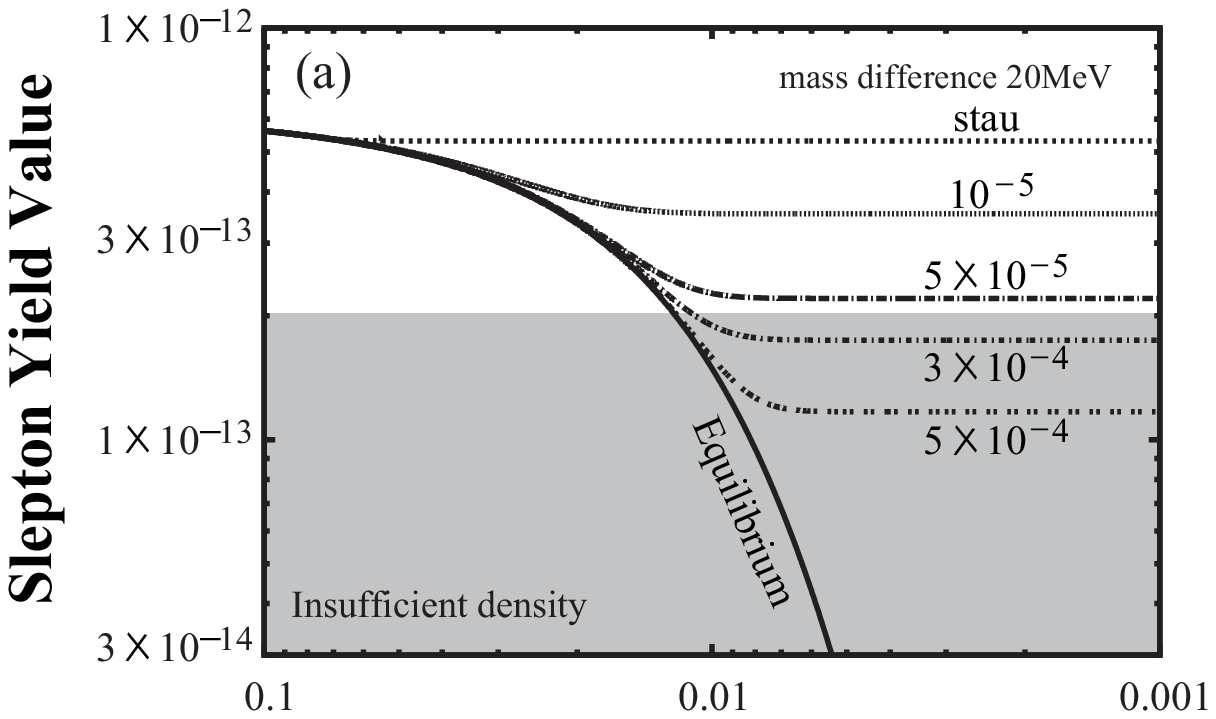}
\\[-4mm]
\includegraphics[width=8.6cm,clip]{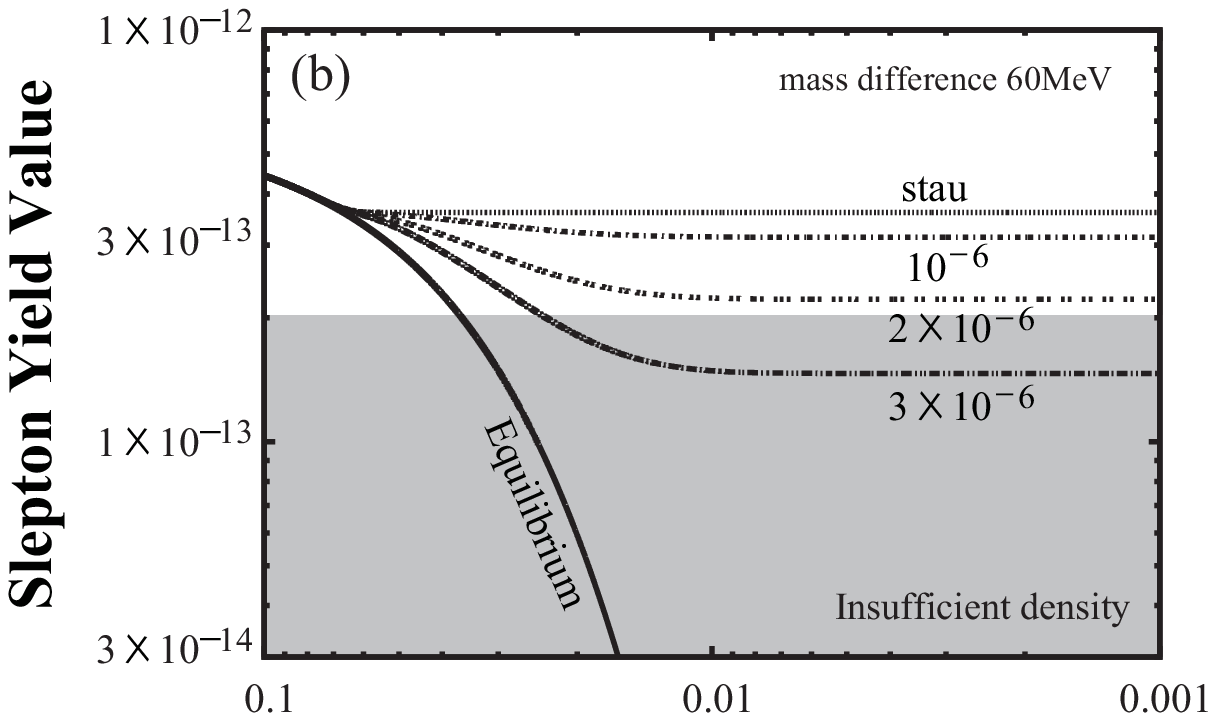}
\\[-4mm]
\includegraphics[width=8.6cm,clip]{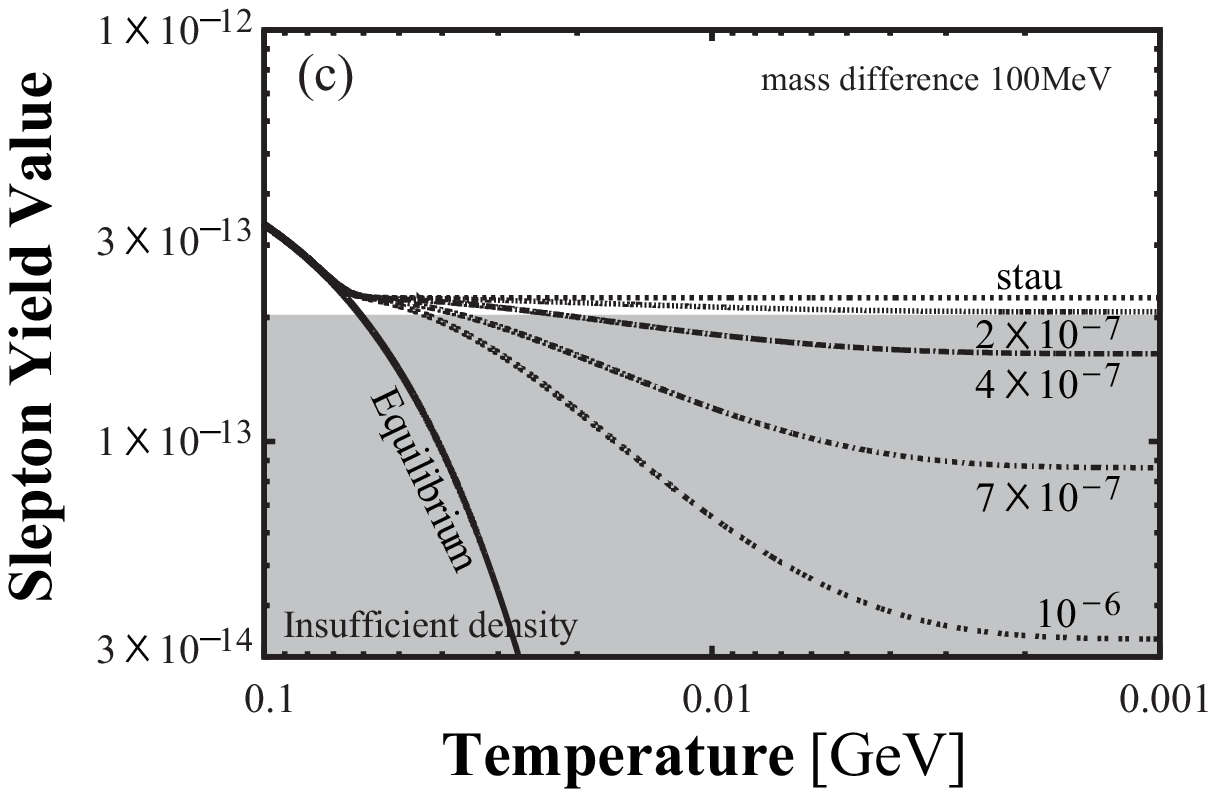}
\caption{Evolution of slepton yield values $Y_{\tilde l}$ for each smuon 
component $c_\mu$, which values are attached on corresponding curve. 
Curves tagged by ``stau" are yield values of pure stau, and curves tagged 
by ``Equilibrium'' are yield values of the slepton in kinetic equilibrium. 
In shaded region, number density of the slepton is insufficient for solving 
the $^7$Li problem. }
\label{fig:cmu}
\end{center}
\end{figure}

Figure \ref{fig:cmu} shows the temperature evolution of the yield
values of the slepton  for each values of $c_\mu$. The value of
$c_\mu$ is shown near corresponding curves.  From top to bottom
panels, the mass differences are taken to be $\delta m = 20$ MeV, $60$
MeV, and $100$ MeV, respectively.
The curves tagged by ``stau'' represent the yield values without the mixing, and 
the curves tagged by ``Equilibrium'' are the yield values of the slepton in kinetic equilibrium. 
In shaded region, number density of the slepton is insufficient to solve 
the $^7$Li problem.
In each panel, we took $c_e = 0$, because bound on $c_e$ is $c_e \lesssim 10^{-10}$ 
and hence contributions of $c_e$ are negligible when the exchange processes 
are still working. 
As is expected, larger value of $c_\mu$ keeps yield values of the slepton to be in 
kinetic equilibrium and leads smaller number densities of the slepton.

The bound on $c_\mu \lesssim 5 \times 10^{-5}$ for $\delta m = 20$ MeV
(Fig.~\ref{fig:cmu}(a)) accurately reproduces our estimation in the
previous section.  For larger $\delta m$, the bounds on $c_\mu$ become
more stringent; $c_\mu \lesssim 2 \times 10^{-6}$ for $\delta m = 60$
MeV (Fig.~\ref{fig:cmu}(b)), and $c_\mu \lesssim 2 \times 10^{-7}$ for
$\delta m = 100$ MeV (Fig.~\ref{fig:cmu}(c)).
This is understood as follows. The cross sections for the processes
(\ref{exc_pro})
are not so dependent on
$\delta m$ for $\delta m < m_\mu$. 
They are more dependent on the muon number density and
hence the temperature. It means that the decoupling temperature
of these processes are almost same as long as $c_\mu$ is same.
Then to have a sufficient yield, these processes must decouple
earlier for larger mass difference. See eq.(\ref{ratio}).
Thus less intergenarational mixing is allowed for larger
mass difference. 

\begin{figure} [!t]
\begin{center}
\includegraphics[width=8.6cm,clip]{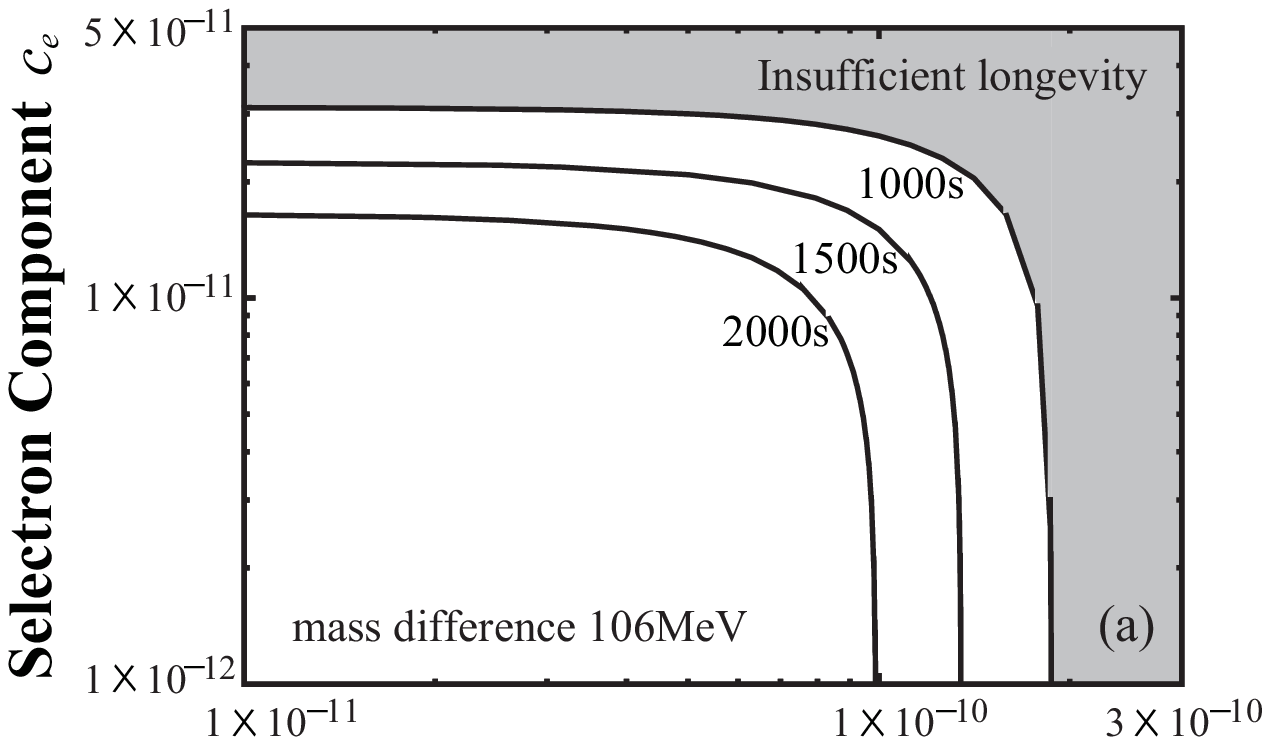}
\\[1mm]
\includegraphics[width=8.6cm,clip]{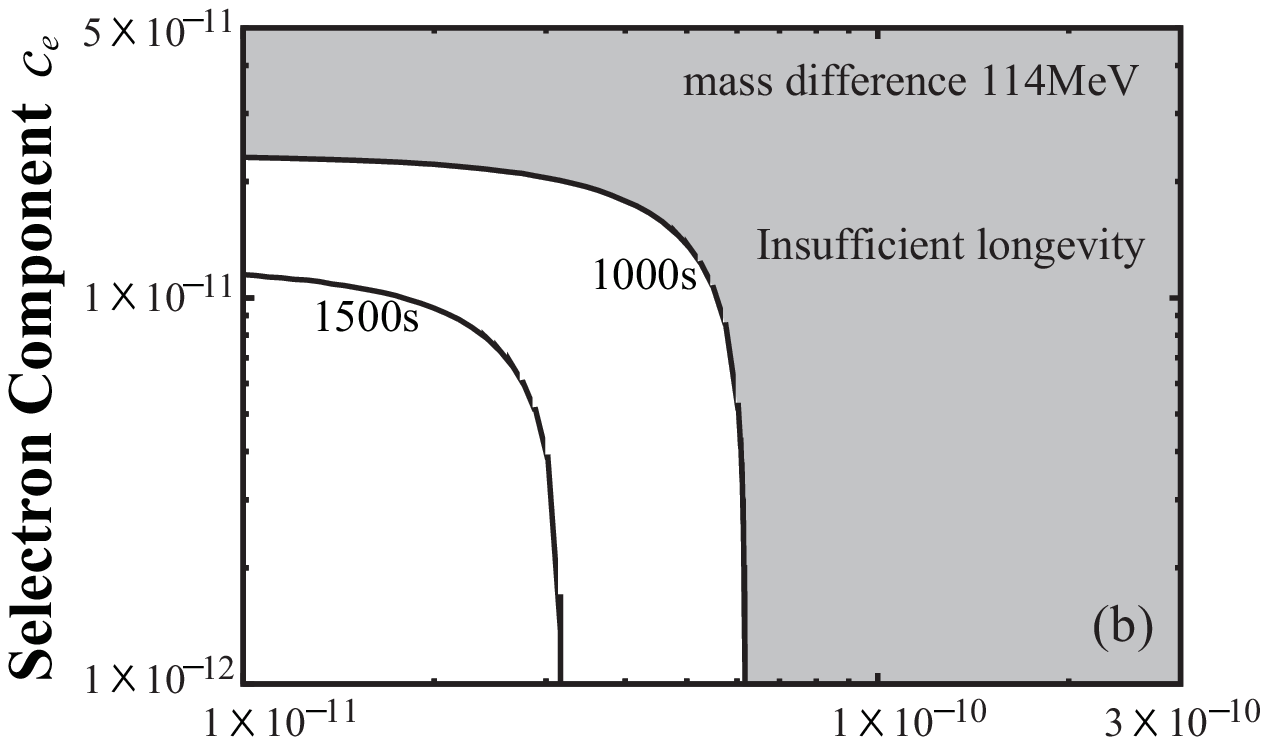}
\\[1mm]
\includegraphics[width=8.6cm,clip]{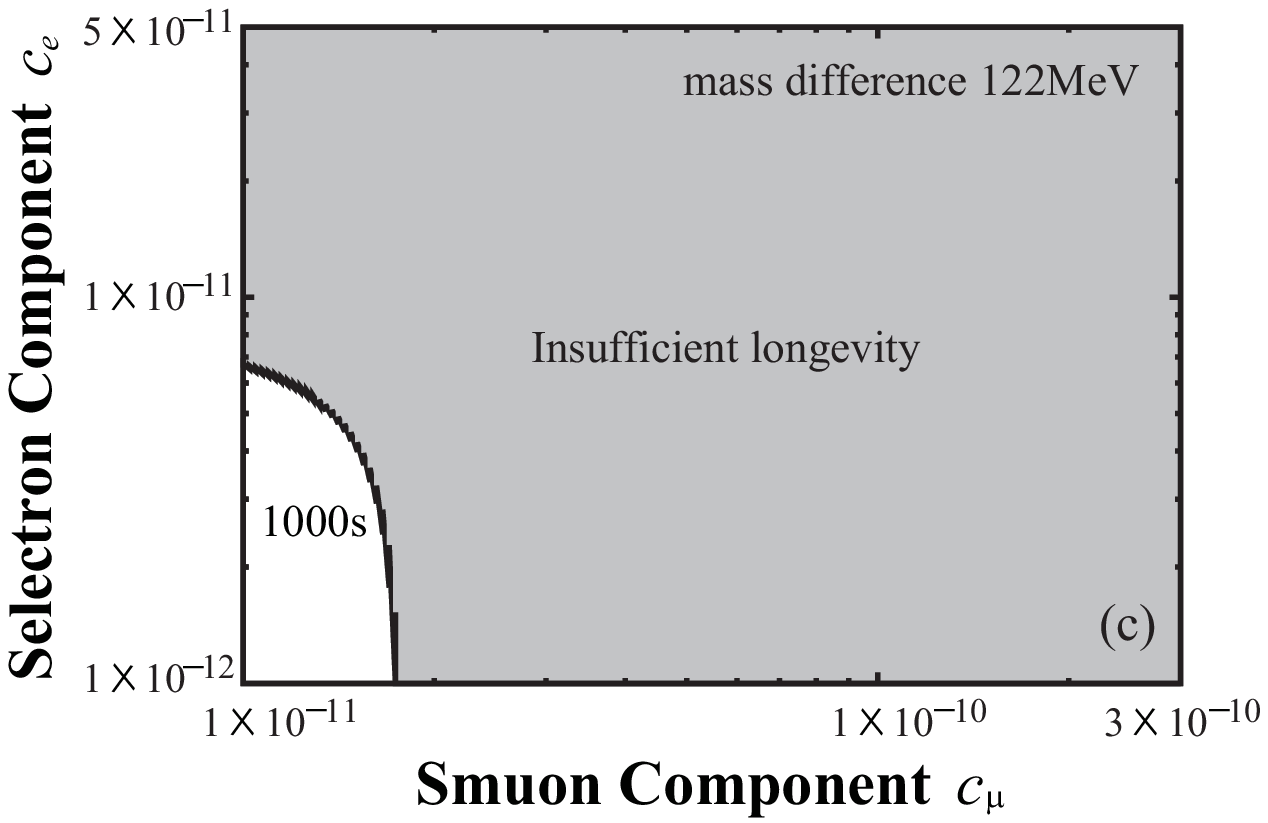}
\caption{Bounds both on $c_\mu$ and $c_e$ for $\delta m \geq m_\mu$. 
Shaded region is disfavored  in light of solving the $^7$Li problem, 
wherein the slepton decays before forming a bound state with $^7$Be and 
$^7$Li.}
\label{fig:cmu_v2}
\end{center}
\end{figure}

In the case of $\delta m \geq m_\mu$, more stringent bounds on $c_\mu$
are   derived from the requirement on lifetimes of the slepton. To
solve the $^7$Li problem,  the bound states of the slepton and $^7$Be
($^7$Li) have to be sufficiently formed.  This formation requires the
lifetime of the slepton to be at least 1000s, preferably  2000s (see
Fig.~1 in Ref.~\cite{Jittoh:2008eq}).
Fig.~\ref{fig:cmu_v2} shows bounds both on $c_\mu$ and $c_e$ for $\delta 
m = 106$MeV, 114MeV, 122MeV, respectively. Values near each curves are the lifetime 
of the slepton. In shaded region, the lifetime of the slepton is shorter than 1000s, 
and hence the parameters in this region are excluded.
In left-side and down-side region in each panel, 4-body decay 
process is dominant because the mixing parameters are so small. 
This result is consistent with the result in Fig.~\ref{fig:life}.  
The upper bounds on $c_\mu$ obtained here are $c_\mu \lesssim 2 \times 10^{-10}$ 
for $\delta m = 106$MeV, which is just above the threshold of the decay of 
the slepton into a muon, and  $c_\mu \lesssim 6 \times 10^{-11}$ for $\delta 
m = 114$MeV, $c_\mu \lesssim 2 \times 10^{-11}$ for $\delta m =
122$MeV.

\begin{figure}[htbp]
    \begin{center}
        \includegraphics[width=8cm,clip]{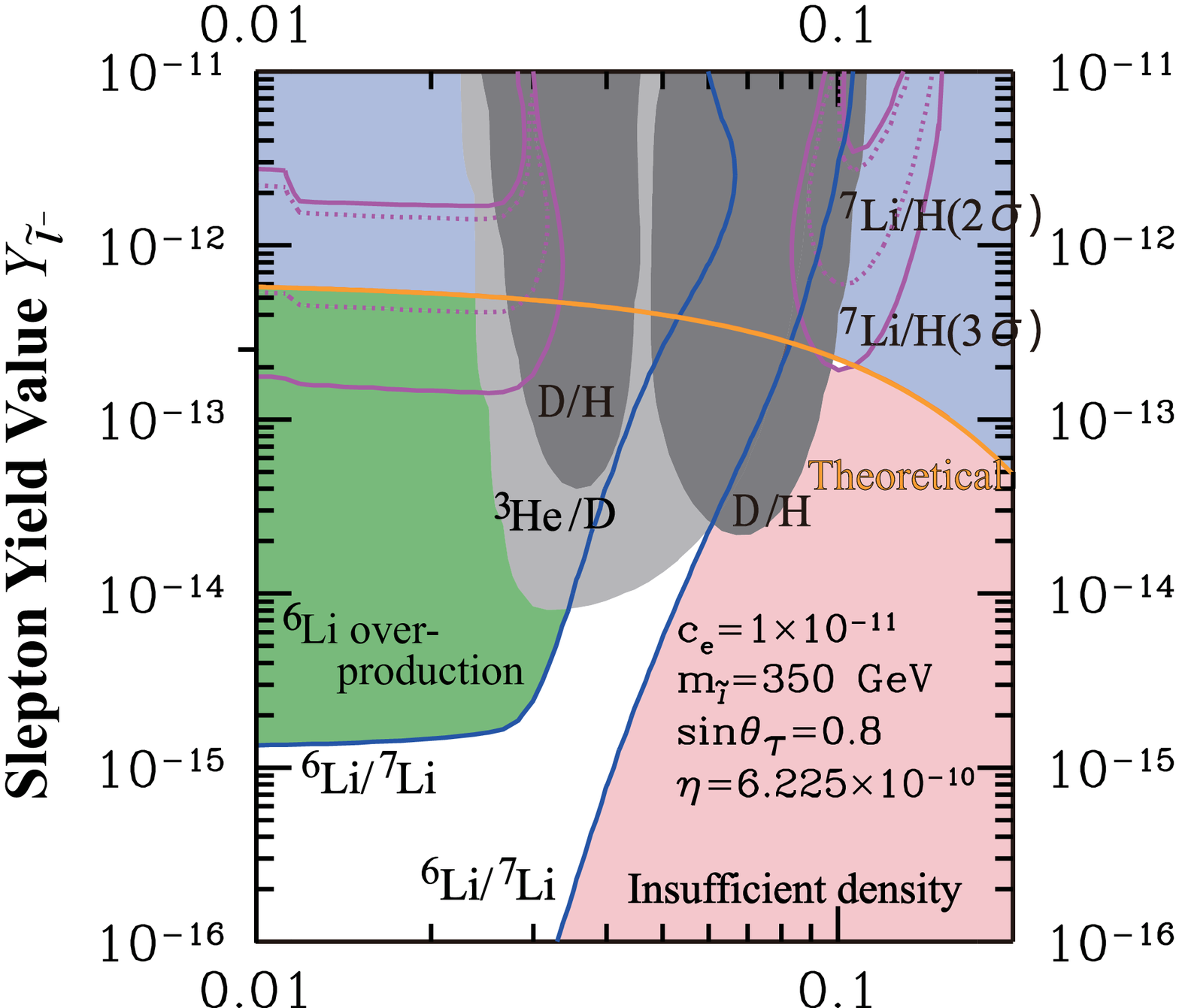}  \\[2mm]
         \label{fig:yield1.E11}
        \includegraphics[width=8cm,clip]{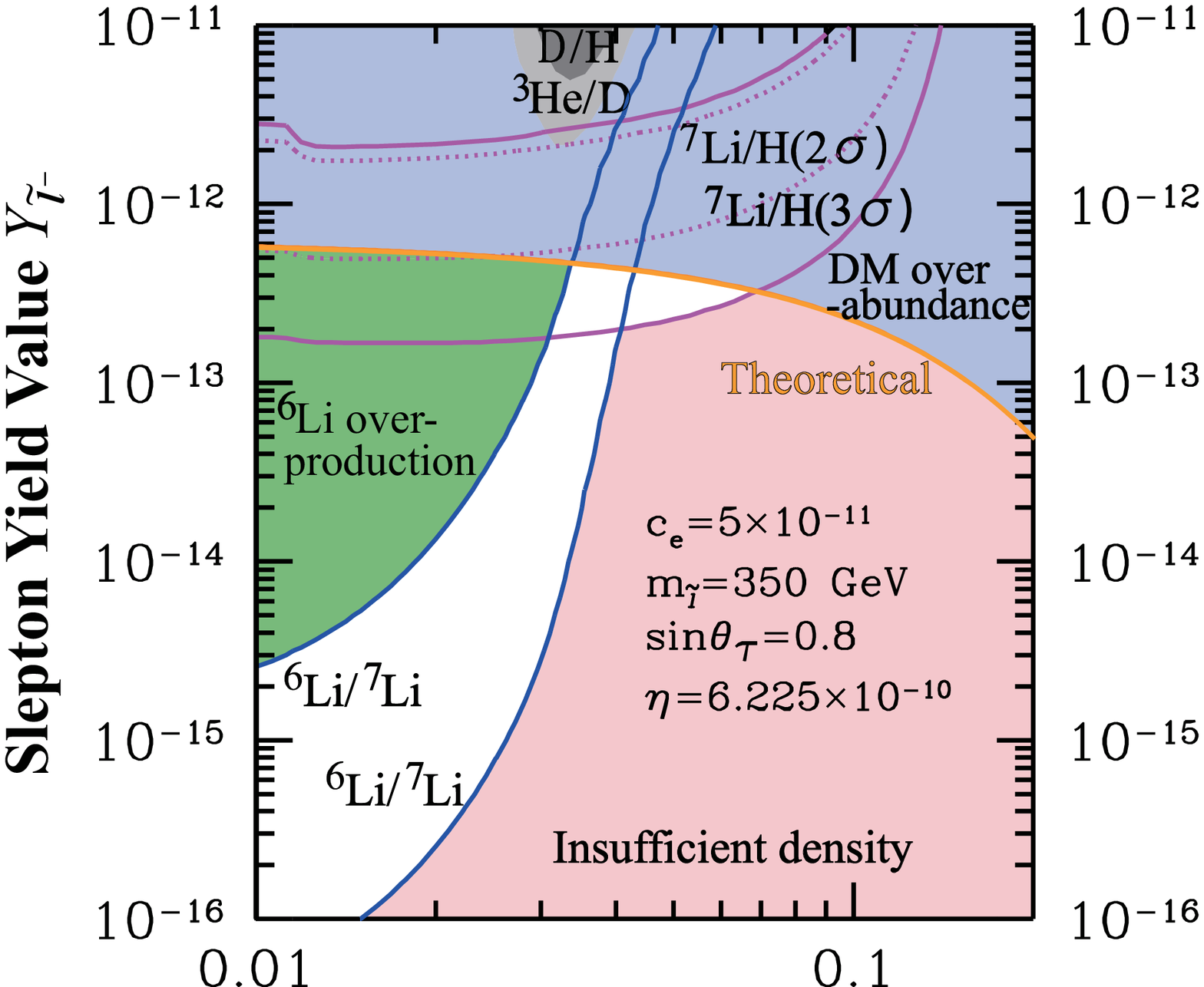}   \\[2mm]
         \label{fig:yield5.E11}
        \includegraphics[width=8cm,clip]{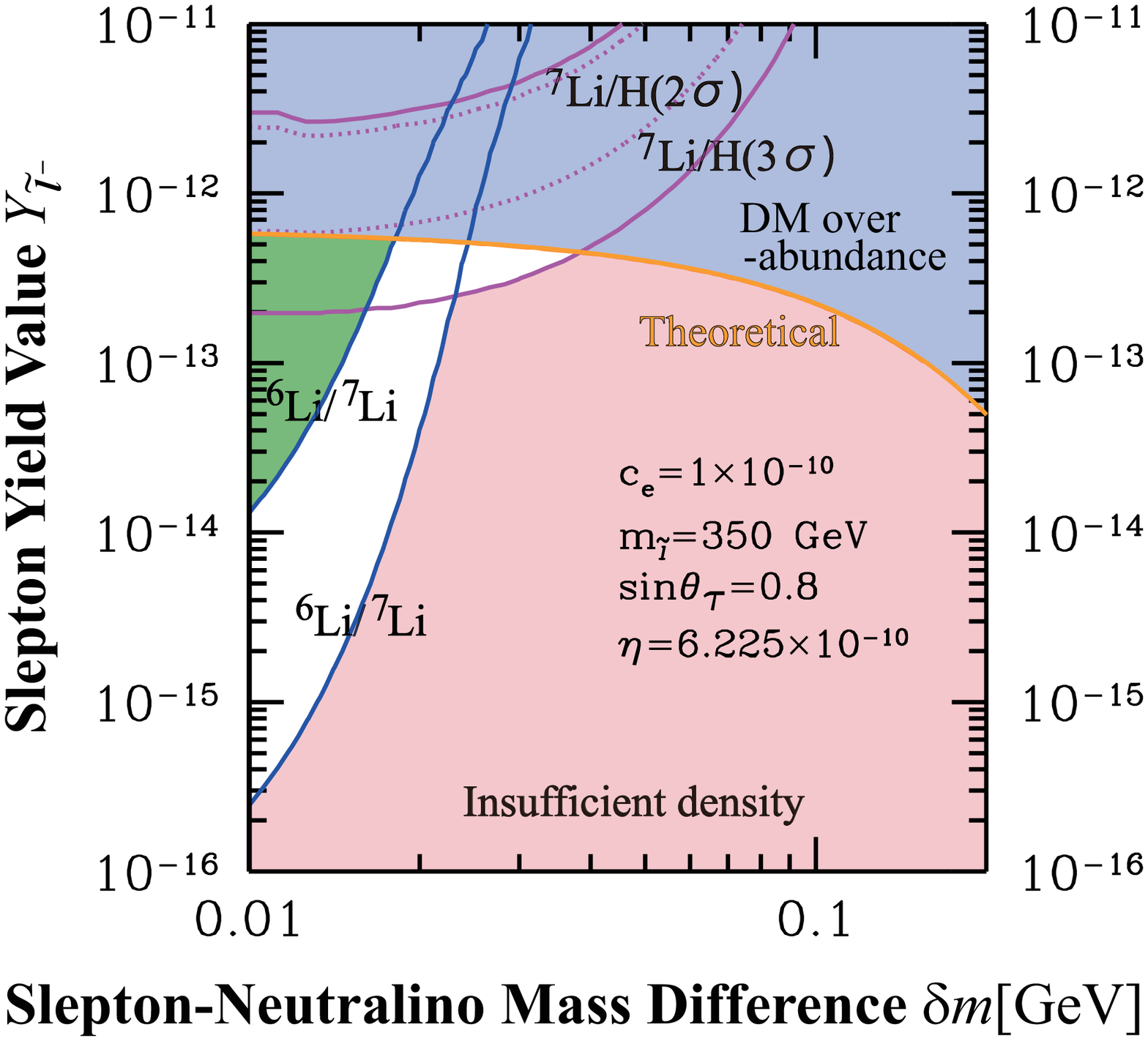}   \\[0mm]
       \caption{ 
       Allowed regions by observational light element abundances in
       $\delta m$--$Y_{\tilde l^{-}}$ planes in cases of
       $c_{e}=10^{-11}$ (top panel), $5 \times 10^{-11}$ (middle
       panel), and $ 10^{-10}$ (bottom panel). The lines of the
       constraints are plotted for D/H, $^{3}$He/D, and
       $^{6}$Li/$^{7}$Li at 2$\sigma$ . An exception is  $^7$Li/H
       whose constraints are denoted by both dotted lines at 2$\sigma$,
       and solid lines at 3$\sigma$. The theoretical curve of the relic
       abundance is also plotted as a thick solid line.}
        \label{fig:yield1.E10}
    \end{center}
\end{figure}

\begin{figure} [!t]
\begin{center}
\includegraphics[width=8.5cm,clip]{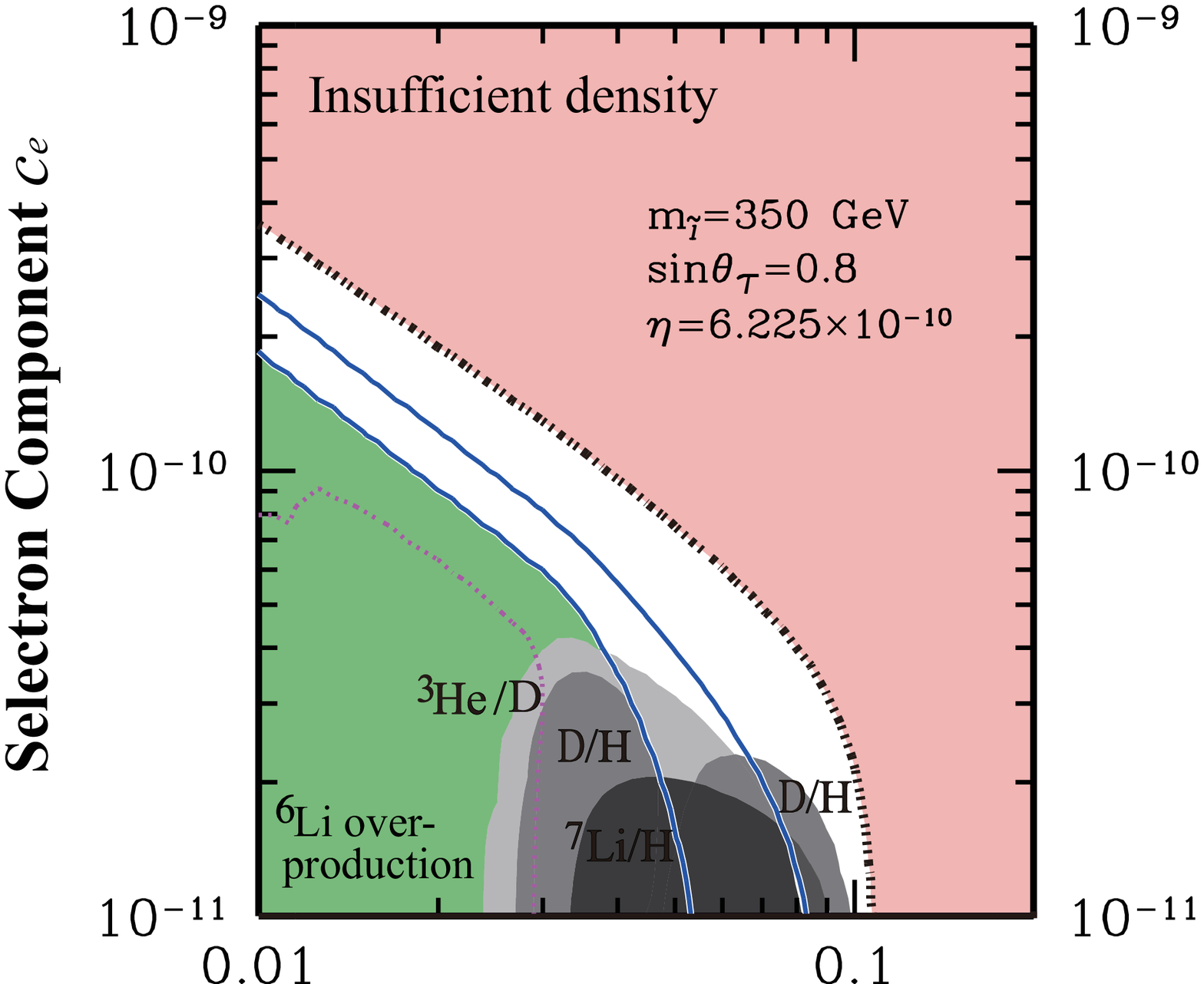}
\\[2.5mm]
\includegraphics[width=8.5cm,clip]{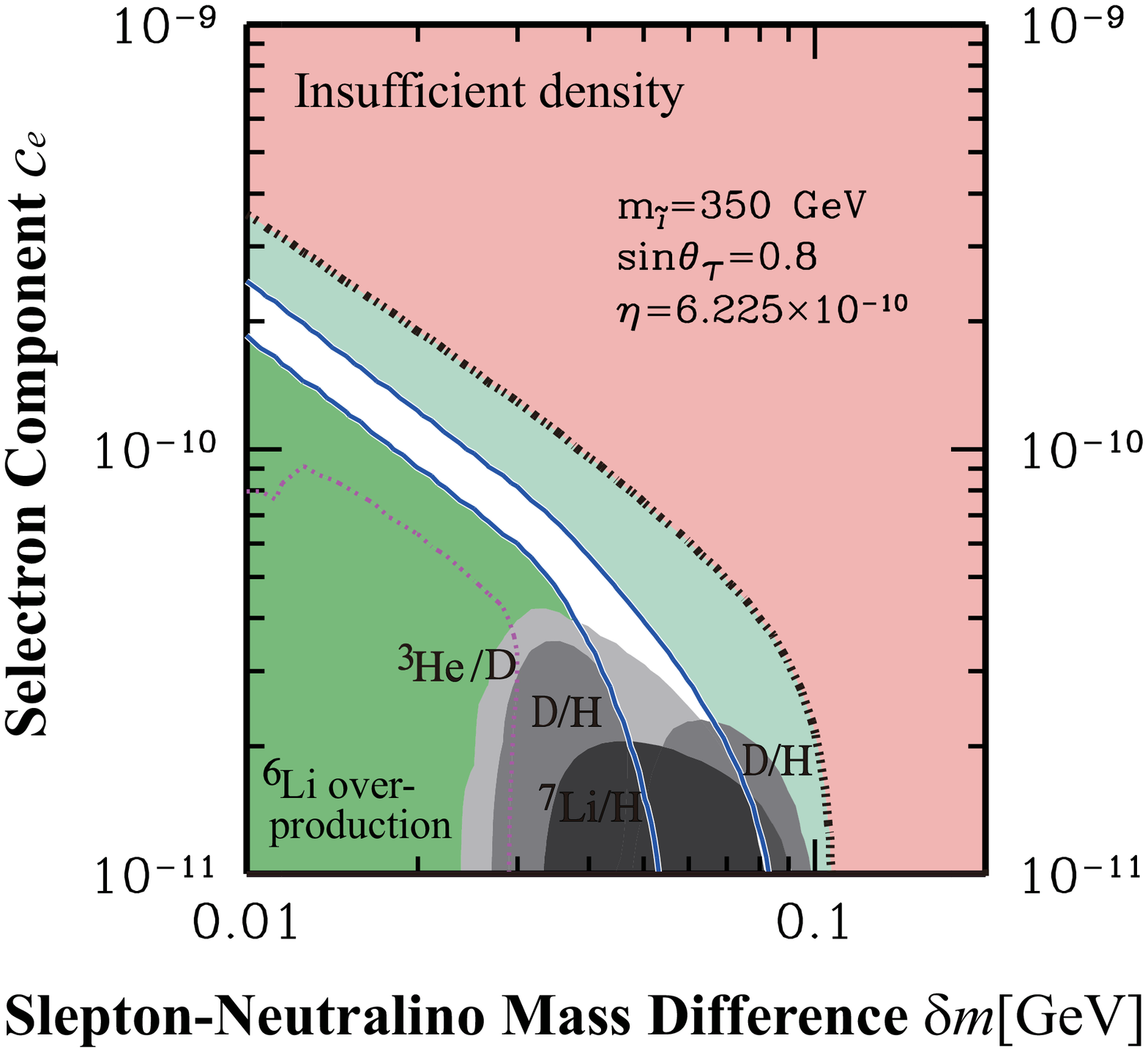}
\caption{
Allowed region in the $\delta m$--$c_e$ plane. Each panel is  plotted
with same parameters. The region on the bottom-left side  of thick
dotted line is free from $^{7}$Li problem  at 3$\sigma$. The regions
between two blue solid lines are allowed for $^{6}$Li/$^{7}$Li. The
$^{4}$He  spallation processes exclude shadowed, light shadowed, and
dark shadowed regions due to over-productions of D/H, $^{3}$He/D,  and
$^{7}$Li/H, respectively. The region on the bottom-left side  of left
blue solid line is excluded by $^{6}$Li over-production. The  white
region in top panel (bottom panel) is the parameter space  which is
consistent with all of observational light element  abundances
including $^{7}$Li (both $^{7}$Li and $^{6}$Li).  }
\label{fig:ce}
\end{center}
\end{figure}

\subsection{Allowed regions by  big-bang nucleosynthesis}


By  numerically  solving  the  Boltzmann  equations,  we  obtain  time
evolutions  of the  bound states  such as  ($^{4}$He  $\tilde l^{-}$),
($^{7}$Li  $\tilde l^{-}$),  and ($^{7}$Be  $\tilde  l^{-}$) including
charge-exchange  reactions~\cite{Kamimura:2008fx,Kohri:2009mi}.   Once
those bound states are  formed, the elements are immediately destroyed
through internal  conversion processes  for $^{7}$Li and  $^{7}$Be,
and spallation   processes   for    $^{4}$He,   induced   by
bound-state
effects.~\cite{Jittoh:2007fr,Bird:2007ge,Jittoh:2008eq,Jittoh:2010wh}.
Then by solving coupled equations  of the  reactions including those
nonstandard  processes, we can  obtain final  abundances of  the light
elements. We adopt  a value of baryon to photon  ratio $\eta = (6.225
\pm 0.170) \times 10^{-10}$ (68\% C.L.)  which was reported by the
WMAP satellite~\cite{Komatsu:2010fb}.

$^{7}$Be  and $^{7}$Li are efficiently destroyed  through 
internal conversion processes ($^{7}$Be $\tilde l^{-}$) $\to$ $^{7}$Li
+ $\nu_{l}+\tilde \chi_{1}^{0}$ with a following standard process $^{7}$Li
+ $p \to ^{4}$He+$^{4}$He, or another subdominant internal conversion
process ($^{7}$Li $\tilde l^{-}$) $\to$ $^{7}$He + $\nu_{l}+\tilde
\chi^{0}$. Then we can compare this theoretical value  with the
observational abundance. It is notable that most of primordial
$^{7}$Li is produced by primordial $^{7}$Be through its electron
capture at a much later time for the adopted value of $\eta$. We have
also included the production process of $^{6}$Li through the
bound-state effect ($^{4}$He$\tilde l^{-}$) + D $\to$ $^{6}$Li +
$\tilde l^{-}$\cite{Pospelov:2006sc,Hamaguchi:2007mp}.

On the other hand, D and $^{3}$He are nonthermally produced by the
$^{4}$He spallation processes ($^{4}$He$\tilde l^{-}$) $\to$ T +$n$ +
$\tilde \chi_{1}^{0}$ + $\nu_l$, ($^{4}$He$\tilde l^{-}$) $\to$ D + 2 $n$
+ $\tilde \chi_{1}^{0}$ + $\nu_l$, ($^{4}$He$\tilde l^{-}$) $\to$  $p$  +
3 $n$ + $\tilde \chi_{1}^{0}$ + $\nu_l$, and subsequent standard processes
for $n$ and $p$~\cite{Jittoh:2010wh}.  Note that $^{7}$Li or $^{7}$Be
is also secondarily produced by those nonthermally-produced energetic
T and $^{3}$He, which may  worsen the $^7$Li problem partly in the
parameter spaces.

In Fig.~\ref{fig:yield1.E10} we plot regions  allowed by observational
light element abundances in $\delta m$--$Y_{\tilde l^{-}}$ planes in
cases of $c_{e}=10^{-11}$ (top panel), $5 \times 10^{-11}$ (middle
panel), and $10^{-10}$ (bottom panel). Lines of the constraints are
plotted for D/H, $^{3}$He/D, and $^{6}$Li/$^{7}$Li at 2$\sigma$.  An
exception is $^7$Li/H whose lines are denoted by both dotted lines at
2$\sigma$, and solid lines at 3$\sigma$. The theoretical curve of the
relic abundance is also plotted as a thick solid line. For simplicity,
here we have assumed $c_{\mu} = 0$.

In Fig.~\ref{fig:ce}, we also plot allowed regions in the $\delta
m$--$c_e$ plane by using the theoretical value of the relic abundance
for the negatively charged slepton at each point of the plane. Each 
panel is plotted with same parameters.

The region on the bottom-left side of thick dotted line is free from
$^{7}$Li problem at 3$\sigma$, and the region on the other side of the
line is excluded  due to insufficient density of the slepton. The
regions between two blue solid lines are allowed for $^{6}$Li/$^{7}$Li 
at 2$\sigma$. The $^{4}$He spallation  processes exclude
shadowed, light shadowed, and dark shadowed regions due to
over-productions of D/H, $^{3}$He/D, and $^{7}$Li/H at 2$\sigma$,
respectively.  The region on the bottom-left side of left blue solid
line is excluded by $^{6}$Li  over-production.  Thus the white region
in top panel (bottom panel) is the parameter  space which is
consistent with all of observational light element abundances
including $^{7}$Li (both $^{7}$Li and $^{6}$Li).

\section{Collider signature}  \label{sec:col} 

\begin{figure} [!t]
\begin{center}
\includegraphics[width=8.5cm,clip]{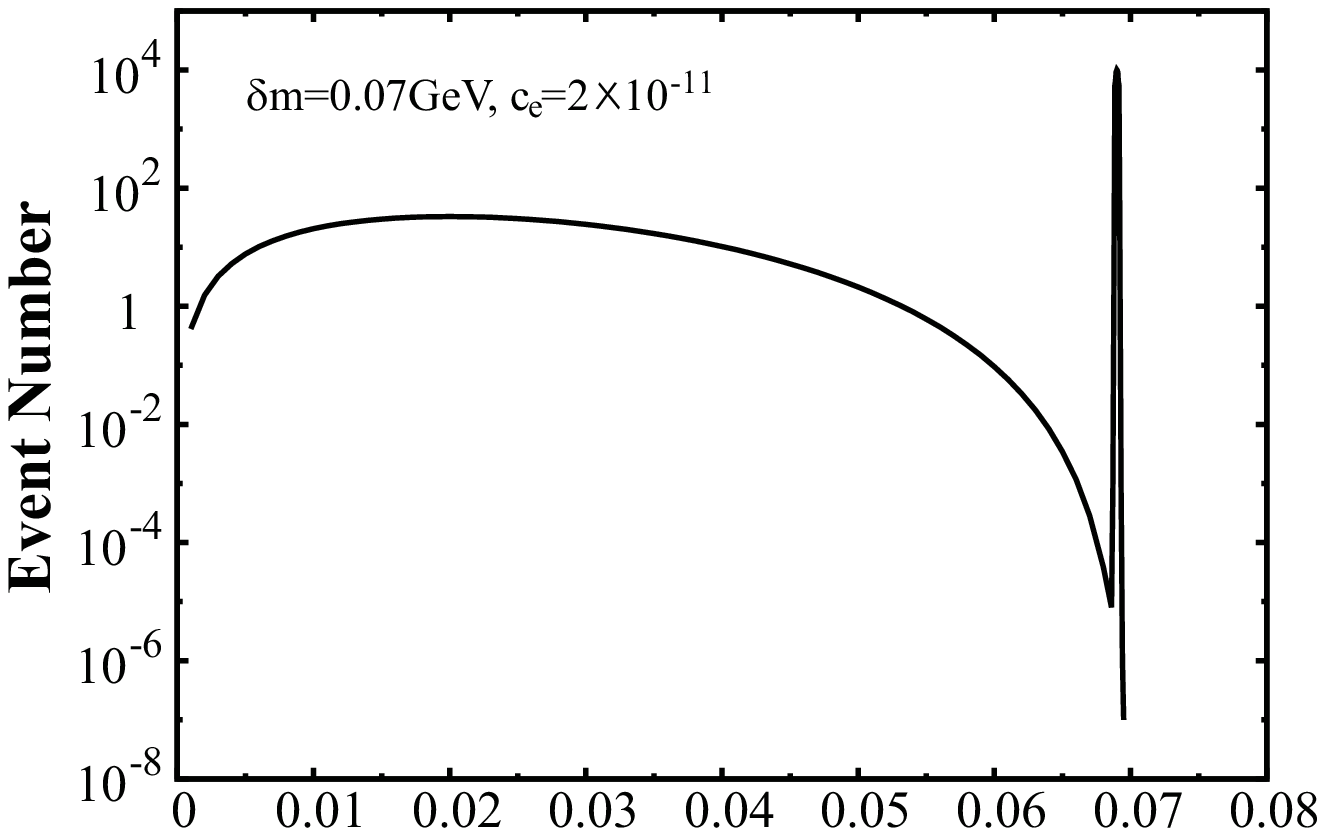}
\\[3mm]
\includegraphics[width=8.5cm,clip]{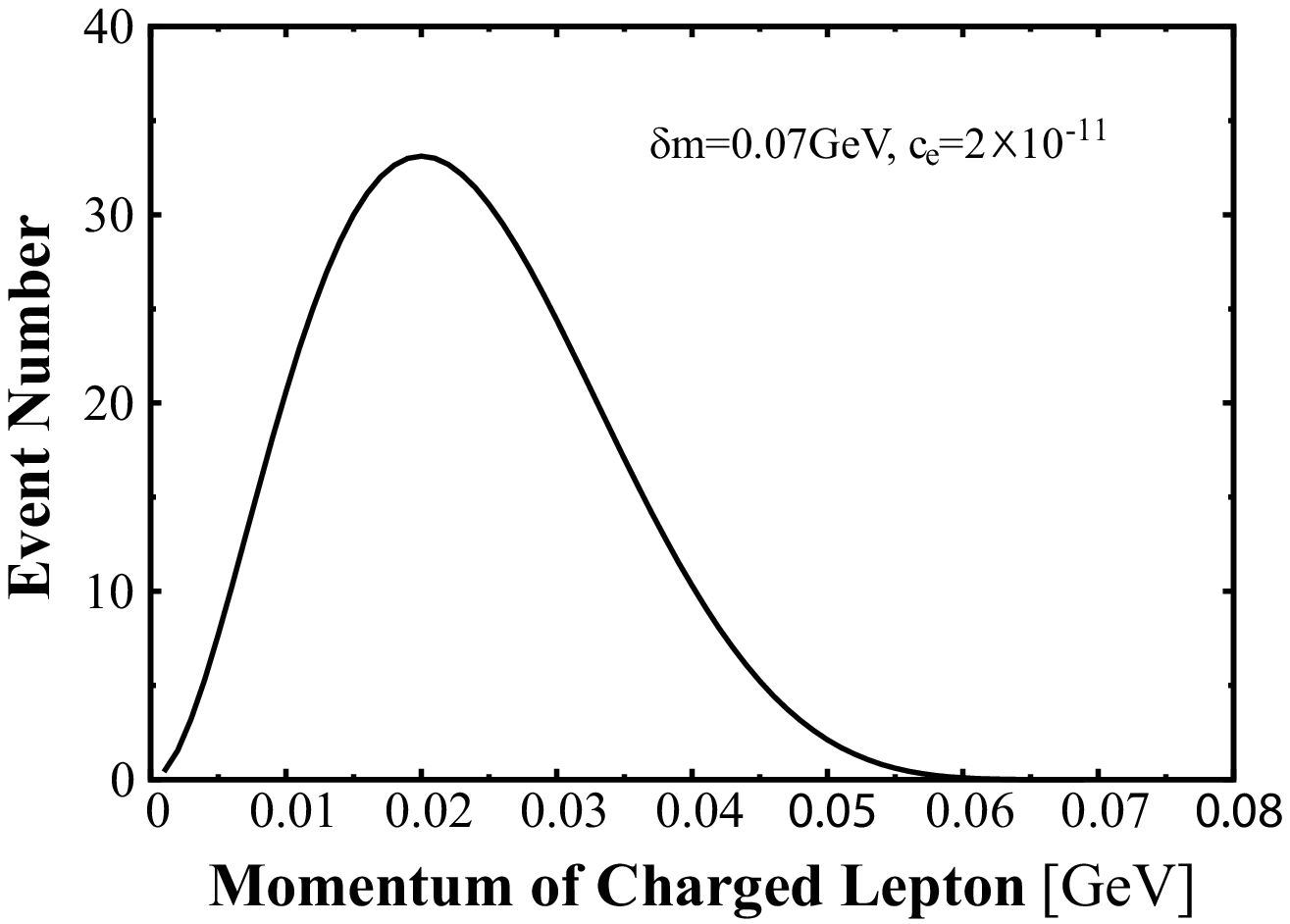}
\caption{Event number distribution of the decay of the long-lived slepton 
as a function of momentum of final state electron. We assumed detector 
resolution to be 1MeV, and normalized number of 2-body decay to $10^4$ 
events in a bin.}
\label{fig:decay_dist}
\end{center}
\end{figure}

Finally, we have a short discussion on collider signatures of the long-lived slepton. 
The long-lived sleptons stopped in a detector decay into 2-body final state ($\tilde l \to 
\tilde \chi_1^0 e$) and 4-body final state  ($\tilde l \to \tilde \chi_1^0 \nu_\tau 
e \bar \nu_e$). From the results in previous section, we can predict the ratio of the number of 
events between these decays.  
One of the way to confirm our scenario is to search for double peaks in  
electron momentum distribution in the decay of the long-lived slepton, and 
count the event number ratio.

Figure \ref{fig:decay_dist} shows expected event distributions of the decay 
as a function of electron momentum. The event distributions of both 2-body and 
4-body decays are plotted in log scale in top panel, and bottom panel plots 
the distribution of only 4-body decay in linear scale.  We took $m_{\tilde l} 
= 350$GeV, $\delta m = 0.07$GeV, $c_e = 2 \times 10^{-11}$, and 
$c_\mu = 0$. This parameter set is found in allowed region (see 
Fig.~\ref{fig:ce}). We assumed detector resolution to be 1MeV, and normalized 
number of 2-body decay to $10^4$ events in a bin.
One can see in the top panel that there is a peak at the momentum of $0.07$GeV.
The peak corresponds to 2-body decay because the momentum is the same value of 
$\delta m$. On the other hand, in the bottom panel, one can see 
that 4-body decay also shows a peak around $0.2$ GeV. This peak is broader 
than that of the 2-body decay. Number of event of 2-body decay is almost 100 times 
more than peak number of 4-body decay. Thus, the event distribution shows a double 
peak structure, and this is a characteristic feature for mass-degenerate scenario.

\section{Summary}  \label{sec:sum} 

We have considered a scenario of the MSSM where the slepton NLSP is
long-lived  due to a small mass difference between the NLSP and  the
neutralino LSP, and studied the effects of the  intergenerational
mixing of sleptons on big-bang  nucleosynthesis.  In this scenario,
the so-called internal conversion processes occurs in the bound
states between the slepton NLSP and light nuclei in BBN. Then, the
$^7$Li and $^6$Li problems can be solved simultaneously when the
slepton NLSP is enough  long-lived and its number density is
sufficient at the era of BBN. We have analyzed  the yield value and
the lifetime of the slepton with the intergenerational mixing as  well
as the relic abundances of the light elements including the internal
conversion  processes in BBN.

In Sec. III, we have calculated the yield value and the lifetime of
the slepton NLSP  and derived upper bounds on the mixing parameters
$c_\mu$.  The upper bounds are obtained by requiring the yield value
and the lifetime to  be $Y_{\tilde{l}} \sim 2 \times 10^{-13}$ and
$\tau_{\tilde{l}} \ge 10^{3}$ sec. to solve the $^7$Li problem.  In the
case of $\delta m < m_\mu$, the upper bounds are given as $c_\mu  \le
5 \times 10^{-5}, ~2 \times 10^{-6}$ and $2 \times 10^{-7}$ for
$\delta m = 20,~60$ and $100$ MeV, respectively.  On the other hand,
in the case of $\delta m > m_\mu$, the upper bounds,  $c_\mu \le 2
\times 10^{-10}, ~6 \times 10^{-11}$ and $2 \times 10^{-11}$ ,  are
given for $\delta m = 106,~114$ and $122$ MeV, respectively.

We have also analyzed the the relic abundances of the light elements
by solving  the Boltzmann equations taking into account 
all the exotic processes listed in Sec.\ref{sec:result}.
We derived an allowed region on $c_e$ by
adopting $2 \sigma$ uncertainties  for D/H,~$^3$He/D and
$^6$Li/$^7$Li, and $2 \sigma$ and $3 \sigma$ uncertainty  for
$^7$Li/H. We found that, assuming $c_\mu=0$, the $^7$Li and $^6$Li
problems can be solved simultaneously in the range of $2 \times
10^{-11} \le  c_e \le 2 \times 10^{-9}$. These results are consistent
with the estimations  given in Sec. II.

In the end, we have discussed the decays of the slepton in a case of
$\delta m  = 0.07$ GeV and $c_e=2 \times 10^{-11}$ as an illustrating
example.  We found that the $2$-body decay shows a sharp peak at the
momentum of  an outgoing electron close to while the $4$-body decay
shows a broad peak  in the momentum distribution. The sharp peak is
found at the momentum equal  to $\delta m$, and the expected number of
events for $2$-body decay is $100$  times larger than that for
$4$-body decay.  These features as well as heavy charged tracks are
unique in the  degenerate-mass scenario. In the case that lifetime of
the long-lived slepton is  $10^{3}$~sec, even present working
detectors can stop the produced long-lived sleptons and
accumulate  $\sim$0.5\%   of them approximately~\cite{Asai:2009ka}. If
exotic charged tracks of long-lived particle are discovered, new
facility may be constructed with higher  accumulation efficiency and
high momentum resolution.  Discovery of unique signatures of the decay
in addition to charged tracks  of long-lived particle will be a strong
evidence of this scenario.

\section*{Acknowledgments}   

This  work was supported in part by the Grant-in-Aid for the Ministry
of Education, Culture, Sports, Science, and Technology, Government  of
Japan, No.~21111006, No.~22244030, No.~23540327(K.K.), No. 23740190
(T.S.), No. 24340044 (J.S.),  and No. 23740208 (M.Y.).


\end{document}